\documentclass[aps,prb,twocolumn,superscriptaddress,showpacs,floatfix,amsmath,amssymb]{revtex4-1}

\newcommand{\beq}[0]{\begin{equation}}
\newcommand{\eeq}[0]{\end{equation}}
\newcommand{\beqa}[0]{\begin{eqnarray}}
\newcommand{\eeqa}[0]{\end{eqnarray}}

\newcommand{\bd}[0]{\begin{description}}
\newcommand{\ed}[0]{\end{description}}

\usepackage{float}
\usepackage{diagbox}
\usepackage{adjustbox}
\usepackage{graphicx,xcolor}
\usepackage{mathtools}
\usepackage{tabularx}
\usepackage{supertabular}

\begin{document}
 
\title{Multifractal study of quasiparticle localization in disordered superconductors}
 
\author{C. W. Moore}
\affiliation{Department of Physics \& Astronomy, Louisiana State University, Baton Rouge, Louisiana 70803, USA}
\author{Ka-Ming Tam}
\affiliation{Department of Physics \& Astronomy, Louisiana State University, Baton Rouge, Louisiana 70803, USA}
\author{Yi Zhang}
\affiliation{Department of Physics \& Astronomy, Louisiana State University, Baton Rouge, Louisiana 70803, USA}
\author{M.\ Jarrell}
\affiliation{Department of Physics \& Astronomy, Louisiana State University, Baton Rouge, Louisiana 70803, USA}

\begin{abstract}

The thermal metal to thermal insulator transition due to random disorder is studied in the context of the symmetries of the Bogoliubov de Gennes Hamiltonian. 
We focus on a three dimensional system with gapless s-wave pairing that possesses time reversal and spin rotational symmetry. The quasiparticle excitations (bogolons) undergo a metal insulator transition as the disorder increases. We determine the critical disorder strength and correlation exponent first by the transfer matrix method (TMM).
We then apply a multifractal finite sized scaling (MFSS) of the bogolon wavefunction obtained from large scale diagonalization of the Hamiltonian and obtain the critical disorder strength and exponent, in agreement with those found by
TMM.


\end{abstract}


\maketitle
 
\section{Introduction}
\label{sec:intro} 

Anderson localization involves the localization of single-particle electronic states in a disordered metal\cite{Anderson1958}. Although this has proved to be a challenging and complex problem\cite{KMreview}, the basic interpretation of the transition is clear: it is a transition from a metallic phase where electrons are able to diffuse and transport over long distances to an insulating phase where this is prevented.  
Anderson localization occurs in normal electronic systems (most famously doped\cite{PaalanenThomas1983} and  amorphous\cite{MottDavis1979} semiconductors). The conducting electronic states are separated from the insulating states by a mobility edge in energy and disorder strength.  Many features of the localization transition 
have been studied and much attention has been paid to two in particular: the multifractality of critical wave functions at the transition and the role played by 
the symmetries of the Hamiltonian 
\cite{Atland1997,Schnyder2008,Wigner1955,Castellani1986}.

The Anderson transition was first and most studied for Hamiltonians of the three Wigner-Dyson\cite{Wigner1955} symmetry classes. The identification of additional symmetry classes (bringing the full number to ten\cite{Atland1997}) has lead
to the study of the effects of Anderson localization beyond the original
three symmetry classes and the additional rich phenomena\cite{Ferdinand2008}. 
In this paper, we consider the question of quasiparticle localization in the Bogoliubov de Gennes class for three dimensions with time reversal and spin rotation symmetry (class CI) which we use to model a dirty superconductor with a finite density of states at the Fermi level.  The excitations of this class are Bogoliubov quasiparticles\cite{Bogoliubov1958} 
(also referred to as bogolons in this paper) with no definite charge as they are a superposition of electron and hole excitations \cite{Beenakker2015}, so this is different from the case of the Anderson model where the excitations have a well defined charge.  In this case, the localization transition is interpreted as localization of bogolons that occurs within the superconducting phase. The two phases are refereed to as a ``thermal metal'' where the bogolons are extended and a ``thermal insulator'' where they are localized\cite{vish1999}.  As mentioned above, the quasiparticles do not transport charge and so there is no Weidemann-Franz law between the thermal and electric transport, but there is still thermal transport and so on the localized side of the transition the system will be thermally insulating and on the extended side it will be thermally metallic \cite{vish1999}.

The idea of multifractality was introduced by Mandlebrot\cite{Mandelbrot1974,Halsey1986} and describes spatial structures that have a complicated distribution and require an infinite number of critical exponents to describe the scaling of their moments.  The multifractal nature of the wavefunction at criticality was realized for Anderson transitions \cite{wegner1980,Castellani1986} and is now recognized as a defining characteristic. A proposed generalization of the multifractal analysis can be used to calculate the critical parameters of the Anderson transition\cite{Rodriguez2010,Rodriguez2011,Ujfalusi2015} which has even been applied to calculations of doped semiconductors\cite{harashima2014}.

In this paper, we apply the generalized multifractal finite size scaling (MFSS)
\cite{Rodriguez2010,Rodriguez2011}  analysis to a simple model of a dirty superconductor. The model Hamiltonian and methods of extracting critical parameters which include transfer matrix method and multifractal analysis are described in Sec.\ref{sec:Theory}. We will demonstrate that the multifractal analysis can be used to extract the critical disorder strength by showing agreement with transfer matrix method calculations and confirms
that this transition falls outside the Wigner-Dyson symmetry class.  Also, we will argue that the multifractal character of the wavefunctions can possibly explain some experimental findings on dirty superconductors, such as the increase in $T_c$ with disorder. These results are presented in Sec.\ref{sec:results} and discussed in Sec.\ref{sec:discussion}.  We conclude in Sec.~\ref{sec:conclusion}

\section{Model and methods}
\label{sec:Theory}

\subsection{Model of Dirty Superconductor}
\label{sec:Hamiltonian}
We study our model of a dirty superconductor within the mean field Bogoliubov-de Gennes approximation,
and so the Hamiltonian is given by
\begin{equation}
H = \sum_{i,j}[t_{i,j}\sum_{\sigma=\uparrow,\downarrow}(c_{i,\sigma}^{\dagger}c_{j,\sigma}+H.c) + \Delta_{i,j} (c_{i,\uparrow}^{\dagger}c_{j,\downarrow}^{\dagger}+H.c.)].
\end{equation}
The annihilation operator for site $i$ with spin $\sigma$ is given by $c_{i,\sigma}$, and similarly for the creation operators. We only consider spin one-half fermions in this study, so $\sigma=\uparrow$ or $\downarrow$.  $t_{i,j}$ and $\Delta_{i,j}$ are the hopping and pairing between site $i$ and $j$ respectively. 

Previous studies of dirty superconductors predominately focused on the pairing with conventional s-wave symmetry with on-site pairing which has a spectral gap at the band center. Without disorder, the spectral function  is given by $E({\bf k}) = \sqrt{\Delta({\bf k})^2 + \epsilon({\bf k})^2}$, and for a cubic lattice $\epsilon({\bf k}) = -2t\sum_{i=x,y,z} \cos(k_{i})$. For the case of conventional s-wave pairing, we have $\Delta({\bf k}) = \Delta_{0} $ a constant.  Since we do not expect for gap formation to be required for multifractal behavior of the wavefunction, we instead focus on a gapless superconductor. A simple choice is one with extended s-wave pairing with the same nodal structure as that of the bare dispersion $\epsilon({\bf k})$ \cite{vish2001}, in which  $\Delta({\bf k}) = \Delta_{0} \sum_{i=x,y,z} \cos(k_{i})$. 

Random disorder is introduced via two independent terms, one for the on-site local potential and the other for the on-site pairing.  Following the convention in Ref.~\onlinecite{vish2001}, the total Hamiltonian may be written as 
\begin{equation}
H = H_{0} + H_{dis},
\end{equation}
\begin{equation}
H_{0} = \sum_{<i,j>}[\frac{1}{\sqrt{2}}\sum_{\sigma=\uparrow,\downarrow}(c_{i,\sigma}^{\dagger}c_{j,\sigma}+H.c) + \frac{1}{\sqrt{2}} (c_{i,\uparrow}^{\dagger}c_{j,\downarrow}^{\dagger}+H.c.)],
\end{equation}
\begin{equation}
H_{dis} = \sum_{i}[\epsilon_{i}\sum_{\sigma=\uparrow,\downarrow}(c_{i,\sigma}^{\dagger}c_{i,\sigma}+H.c) + \Delta_{i} (c_{i,\uparrow}^{\dagger}c_{i,\downarrow}^{\dagger}+H.c.)].
\end{equation}
The disorder in onsite potential and onsite pairing is assumed to be uniformly distributed from $-W$ to $W$, and so
$P(\epsilon_{i})=P(\Delta_{i})=1/2W ~ \forall ~ -W < \epsilon_{i},\Delta_{i} < W$.
The Hamiltonian possesses time reversal symmetry, spin 
rotation symmetry and particle-hole symmetry which dictates that eigenstates always come in pairs with energy $E$ and $-E$.
These symmetries put the Hamiltonian into the CI class \cite{Atland1997}.

\subsection{Transfer Matrix Method}
\label{sec:TMM}

We first locate the critical point of the model and its localization length exponent using the transfer matrix method. The three dimensional system has a width and height equal to $M$ for each slice of a $N$-slice cuboid, forming a ``bar''
of length $N$. The Hamiltonian can be decomposed into the form 
\begin{equation}
H = \sum_{i} H_{i} + \sum_{i} (H_{i,i+1}+H.c.),
\end{equation}
where $H_{i}$ describes the Hamiltonian for slice $i$ and $H_{i,i+1}$ is the coupling terms between the $i$ and $i+1$ slices. The Schr\"odinger equation can be written in
the form  
\beq
{H}_{n,n+1} c_{n+1} = (E - { H}_n) c_n - { H}_{n, n-1} c_{n-1}  
\label{andSchro} 
\eeq
where $c_i$ is the $M^{2}$ components wavefunction of the slice $i$. 
We introduce the transfer matrix
\beq
T_i = \begin{bmatrix} H_{i,i+1}^{-1} (E-H_{i}) & -H_{i,i+1}^{-1} H_{i-1,i} \\ 1 & 0 
\end{bmatrix}
\label{transferMatrix}
\eeq
and Eq.\ref{andSchro} can be interpreted as the iteration of 
\begin{equation}
\left[  \begin{array}{c} c_{i+1}\\ c_{i} \end{array} \right] =  T_i \times \left[ \begin{array}{c} c_{i} \\ c_{i-1} \end{array} \right].
\label{tmm_eq}
\end{equation}
The goal of the transfer matrix method is to calculate the localization length, $\lambda_{M}(E)$, from the product of $N$ transfer matricies
\begin{equation}
\tau_{N} \equiv \prod_{i=1}^{N} T_{i}.
\label{Qn}
\end{equation}
The Lyapunov exponents of the matrix $\tau_{N}$ is given by the logarithm of its eigenvalues. The smallest exponent corresponds to the slowest exponential decay of the wavefunction and thus can be identified as corresponding to the localization length, $\lambda_{M}(E)$. The
localization length is computed by repeated multiplication of $T_{i}$, but since the multiplication of matrices is numerically unstable periodic reorthogonalization is needed in the numerical implementation\cite{Kramer2010}.  
We use a QR decomposition
for reorthogonalization implemented by LAPACK\cite{LAPACK}, 
and so at the $s$ 
reorthogonalization step the matrix (corresponding to some 
intermediate $L$'th multiplication in calculating Eq.\ref{Qn})
the matrix is decomposed
\beq
\tau_L = Q R
\eeq
where $R$ is an upper triangular matrix and the Lyapunov exponents
$\gamma_s$ are
calculated as 
\beq
\gamma_s = \gamma_{s-1} + \log b_s 
\eeq
where $b_s$ are the $2M^2$ diagonal elements of $R$ for the $s$
renormalization step.  The multiplication of transfer matrices is then
continued with the $Q$ matrix.
The slowest decaying exponent ($\gamma_\ell$) is used to compute the localization
length $\lambda_M(E)=1/\gamma_\ell$ for a given width $M$ and energy $E$.

The localization length is then used to calculate the 
the Kramer-Mackinnon\cite{MacKinnonKramer1983} scaling parameter $\Lambda_M(E) = \lambda_M(E)/M$ which is expected to scale as 
\beq
\Lambda_M = \frac{\lambda_M}{M} = f\left(\frac{M}{\xi}\right),
\eeq
where $\xi \propto |W-W_c|^{-\nu}$.  The scaling function $f$ is Taylor expanded about the critical point $W_c$ and the critical parameters $W_c$ and $\nu$ enter as fitting parameters and 
so can be determined by a least-squares minimization.

\subsection{Multifractal Analysis}
\label{Sec:MFanalysis}
We consider the multifractal properties of the bogolon wave-function
$|\psi_i|^2 = |u_i|^2 + |v_i|^2$ for a three dimensional simple cubic lattice of linear size $L$.  The method is based on the study of Anderson models in Wigner-Dyson class. \cite{Rodriguez2010,Rodriguez2011,Ujfalusi2015} This cubic wavefunction is  partitioned into boxes of linear size $\ell$.  We introduce the quantity $\lambda=\ell / L$ and so we have $N_b = \lambda^{-d}$ as the number of boxes where $d$ is the dimensionality of the system.  In this paper, we shall 
only consider $d=3$.
We introduce the ``coarse grained'' box measure
\beq
\mu_{b(\ell)} = \sum_{i\in b(\ell)} |\psi_i|^2
\label{boxSum}
\eeq
where $b(\ell)$ indexes the $N_b$ boxes 
for a given box size $\ell$.  We introduce for convenience\cite{Rodriguez2011} the quantity
\beq
\alpha \equiv \frac{\log \mu}{\log \lambda}
\label{alpha}
\eeq
to work with instead of directly with the box measures given in Eq.\ref{boxSum}.
Multifractility implies that the number of boxes that correspond 
to a given $\alpha$ (we denote as $N(\alpha)$) must scale as
\beq
N(\alpha) \sim \lambda^{-f(\alpha)}
\label{NalphaDef}
\eeq
where $f(\alpha)$ is some fractal dimension that depends on $\alpha$.  For the
case where $|\psi|^2$ are distributed uniformly in space, one would expect
there to be only a singular $\alpha$ and from the definition of $\lambda$ above
$f(\alpha) = d$.  However, for finite $\lambda$ a narrow distribution 
peaked around $f(\alpha) = d$ would be expected and so the above 
Eq.\ref{NalphaDef} is only defined in the limit $\lambda \rightarrow 0$.
The fact that there exists an $\alpha$ dependent \emph{spectrum} $f(\alpha)$
characterizes a system as being multifractal\cite{Nakayama2003}.

We will want to consider the $q$-dependent moments of the distribution
of $\alpha$ or $\alpha(q)$.  We first introduce 
the generalized inverse participation ratios
for the coarse grained distributions $P(\mu_{b(\ell)})$ as
\beq
 R_q =  \sum_{b(\ell)}^{N_b} \left( \mu_{b(\ell)} \right)^q
\eeq
and assume (similarly to Eq.\ref{NalphaDef}) that the moments of the distribution of each box measure scale by the $q$ dependent exponents $\tau(q)$ or
\beq
\langle R_q \rangle  \sim \lambda^{\tau(q)}
\label{tauQdef}
\eeq
where $\langle \cdot \cdot \cdot \rangle$ denotes an ensemble 
average.
It can be shown\cite{Nakayama2003} 
that  $f(\alpha)$ and $\tau(q)$ can be related by a Legendre transform 
\beq
f(\alpha) = - \tau(q) + q \alpha \, ,
\eeq
where
\beq
\alpha(q) = \frac{d \tau(q)}{d q}.
\label{dq}
\eeq
Carrying out the differentiation in Eq.\ref{dq} and using the 
definition of $\tau(q)$ in Eq.\ref{tauQdef} leads to the 
expression
\beq
\alpha(q) = \lim_{\lambda \rightarrow 0}
\frac{\langle S_q \rangle }{\log \lambda \langle R_q \rangle}
\label{alphaqForm}
\eeq
where
\beq
S_q = \sum_k^{N_b} \mu_k^q \log \mu_k.
\eeq

As defined above, the multifractal exponents are only strictly defined in the limit of infinite system size ($\lambda \rightarrow 0$ as mentioned above) 
and at the critical point.  However, they can be defined for fixed $\lambda$ which we denote with a tilde as
\beq
\tilde{\alpha}_q = \frac{\langle S_q \rangle}
{\log \lambda \langle R_q \rangle}.
\label{alphaqForm}
\eeq
The error in $\tilde{\alpha}_q$, $\sigma_{\tilde{\alpha}}$, 
is then estimated from standard propagation of uncertainty
\begin{equation*}
\left(\frac{ \sigma_{\tilde{\alpha}} }{ \tilde{\alpha} }\right)^2 = 
\left(\frac{\sigma_{\langle S_q \rangle }}{\langle S_q \rangle }\right)^2 +
\left(\frac{\sigma_{\langle R_q \rangle }}{\langle R_q \rangle }\right)^2 - 
2 \left(\frac{\sigma_{\langle R_q \rangle \langle S_q \rangle}}{
\langle R_q \rangle  \langle S_q \rangle }\right)^2 
\end{equation*}
where the covariance term $\sigma_{\langle S_q \rangle 
\langle R_q \rangle}$ is kept to
account for correlations as $R_q$ and $S_q$ are computed from the
same data set.

The quantity $\tilde{\alpha}_q$  scales according to standard one parameter scaling for fixed $\lambda$ in
a relevant
($\rho$) and an irrelevant ($\eta$) scaling variable or \cite{Rodriguez2010,Rodriguez2011}
\beq
\tilde{\alpha}_q(W,L) = G(\rho L^{1/\nu},\eta L^{-|y|}). 
\eeq
We expand the scaling function to first order in the irrelevant operator $\eta$
\beq
\tilde{\alpha}_q(W,L) = G^{(0)}(\rho L^{1/\nu}) + \eta L^{-|y|} G^{(1)}(\rho L^{1/\nu}),
\label{alphaScalingAnsatz}
\eeq
where the sub-leading term is characterized by $\eta, y$, and $G^{(1)}$. The function $G^{(s)}$ (where $s=0,1$ from above) is expanded as a Taylor series
\beq
G^{(s)}(L^{1/\nu}) = \sum_{k=0}^{n_s} a_{sk} \rho ^k L^{k/\nu}.
\label{Geq}
\eeq
The scaling fields $\rho$ and $\eta$ are likewise expanded in terms of $w=(W-W_c)/W_c$
as
\beq
\rho(w) = w + \sum_{m=2}^{m_\rho} b_m w^m
\label{rhoeq}
\eeq
and
\beq 
\eta (w) = 1 + \sum_{m=1}^{m_\eta} c_m w^m.
\label{etaeq}
\eeq 
The critical parameters ($W_c$, $\nu$) and the irrelevant scaling exponent $y$ are determined by fitting the data for $\tilde{\alpha}_q(W,L)$ to Eq.\ref{alphaScalingAnsatz}.  In addition, we have $n_0 + n_1 + m_\rho + m_\eta$ Taylor expansion parameters. 
The correlation length is $\xi = |\rho(\omega)|^{-\nu}$ and so the scaled $\tilde{\alpha}_q(W,L)$ data (which we denote as 
$\tilde{\alpha}^{\rm corr}_q)$ 
 collapses onto two branches
\beq
\tilde{\alpha}^{\rm corr}_q = G_{q}^{(0)} (\pm (L / \xi)^{1/\nu})
\label{scalingCollapse}
\eeq


\section{Results}\label{sec:results}

\begin{figure}[h!]
 \includegraphics[trim = 0mm 0mm 0mm 0mm,width=1\columnwidth,clip=true]{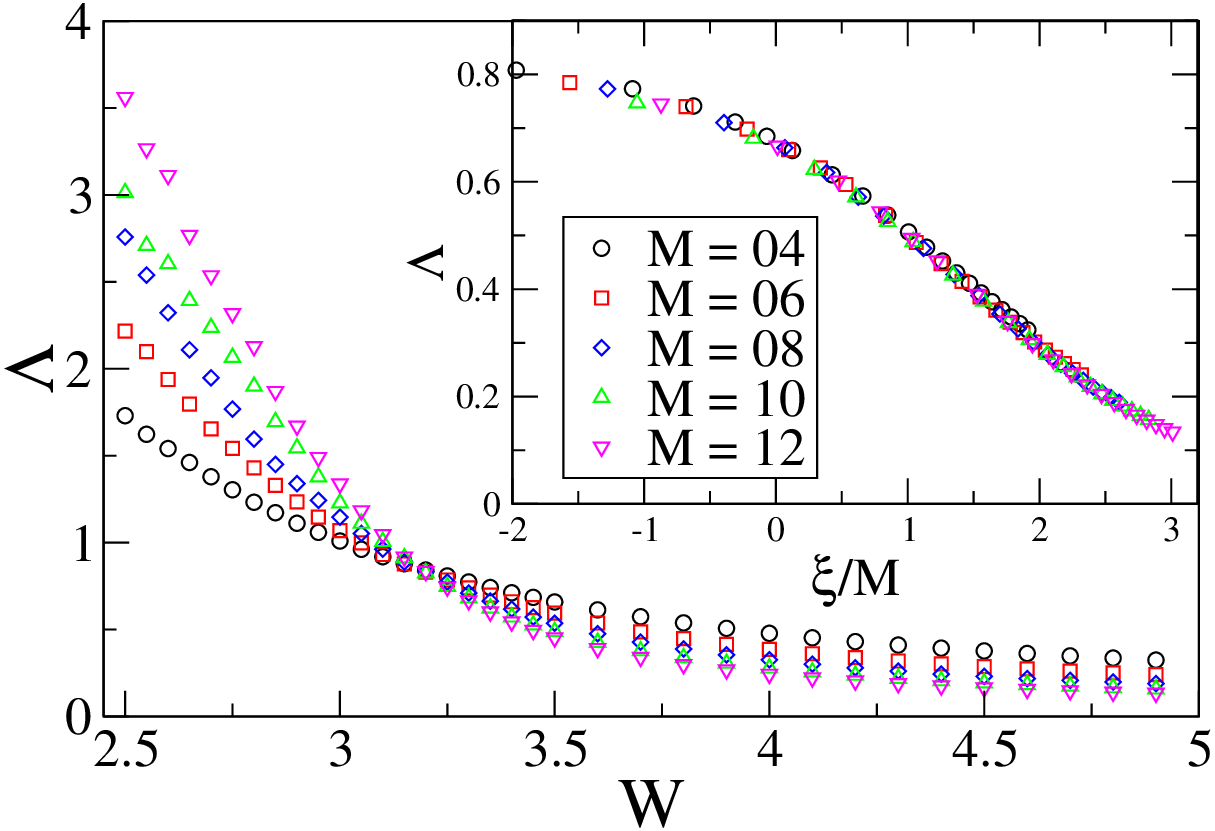}
 \caption{Kramer-Mackinnon scaling parameter as a function of disorder strength calculated with the transfer matrix method
 for a bar of length $N=20000$, $E=0$ and a QR reorthogonalization
 is performed after every $5$ multiplications.
 Note the crossing indicating a
 critical disorder strength around $W=3.2$. When the finite size scaling is performed as described in \ref{sec:TMM}
the data collapses as is shown in the inset.  A bootstrap re-sampling is performed to generate $100$ data sets 
to estimate
the fitting parameters yielding $W_c = 3.212 \pm 0.008$ and a critical exponent of $\nu = 1.01 \pm 0.05$. }
 \label{fig:KMscaling}
\end{figure}

We employ the transfer matrix method to find the critical disorder strength by performing a finite size scaling analysis as shown in Fig.\ref{fig:KMscaling}.  We will compare this result with that predicted by multifractal analysis of the bogolon wavefunction.  The fitting is performed using the SciPy package which acts as a wrapper to MINPACK to perform the least squares minimization \cite{scipy_ref,minpack_ref}.  The fitting range used in Fig.\ref{fig:KMscaling} is determined by performing multiple fits and choosing the one that approximately provides the minimum for the sum of squares.  This range is then used for $100$ bootstrapped resamples of the data to estimate the error bars.  Note however that there can still be error in choosing the fitting range so the error bars are most likely under-estimated.  The calculation was performed for $E=0$ as were are interested
in only the lowest energy excitations which will also be the focus in the following
multifractal analysis.

\begin{figure}[h!]
 \includegraphics[trim = 0mm 0mm 0mm 0mm,width=1\columnwidth,clip=true]{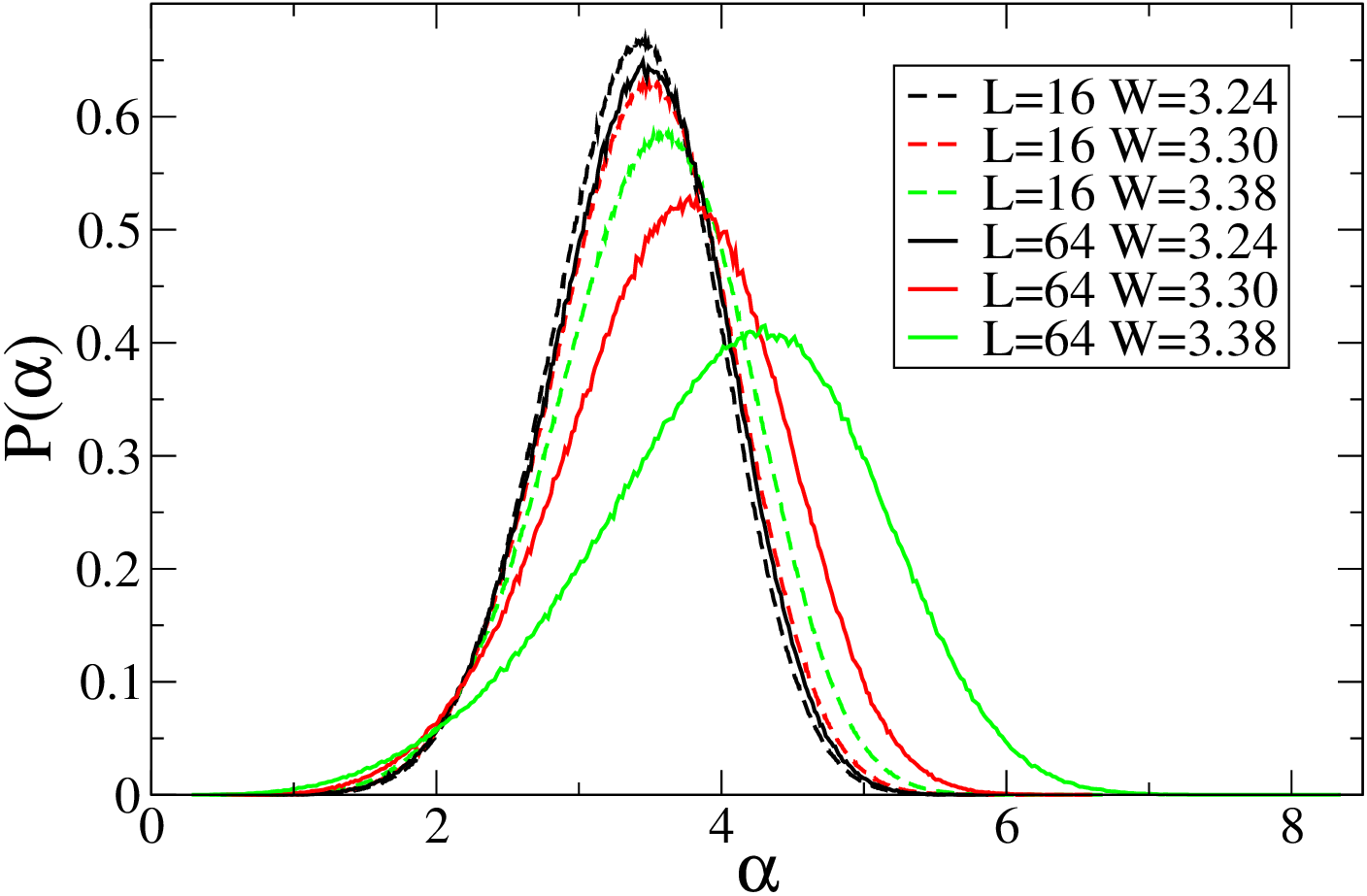}
 \caption{Distribution of the quantity $\alpha$ (defined in Eq.\ref{alpha}) for a finite value of $\lambda = 1/8$ for various system sizes 
 and two disorder strengths.
  The behavior
 of the distributions as a function of $L$ motivates the application of the multifractal analysis in the Ref. \onlinecite{Rodriguez2011} as when the 
 transition is approached ($\sim 3.2$) the distributions become more
 scale invariant (not depending on system size). }
 \label{fig:Alpha0Distribution}
\end{figure}

For the multifractal analysis of the bogolon wavefunctions, we use the JADAMALU package which implements a Jacobi-Davidson method with 
preconditioning\cite{JADAMALU,JADAMALU_code} to 
diagonalize the Hamiltonian. In contrast to that of the 
conventional Anderson model, the disorder terms for the present model appear
in the off-diagonal elements. This poses as a challenge for attaining convergence by the iterative algorithm, both in term of the memory storage and floating point operation. Therefore the accessible system sizes are 
limited in comparison to that of the models with diagonal disorder terms. \cite{Rodriguez2010,Rodriguez2011} Table \ref{paramTable} lists the number of realizations generated for different system size and disorder strength. 
We keep only one state from each realization with the closest eigenvalue (and associated eigenvector) to zero. This is to prevent correlations in wavefunctions that come from the same realization of disorder.  
The wave function can then be coarse grained (as described in Sec.\ref{Sec:MFanalysis}) and the distribution of $\alpha$ 
is plotted in 
in Fig.\ref{fig:Alpha0Distribution}.  

\begin{figure}[h!]
 \includegraphics[trim = 0mm 0mm 0mm 0mm,width=1\columnwidth,clip=true]{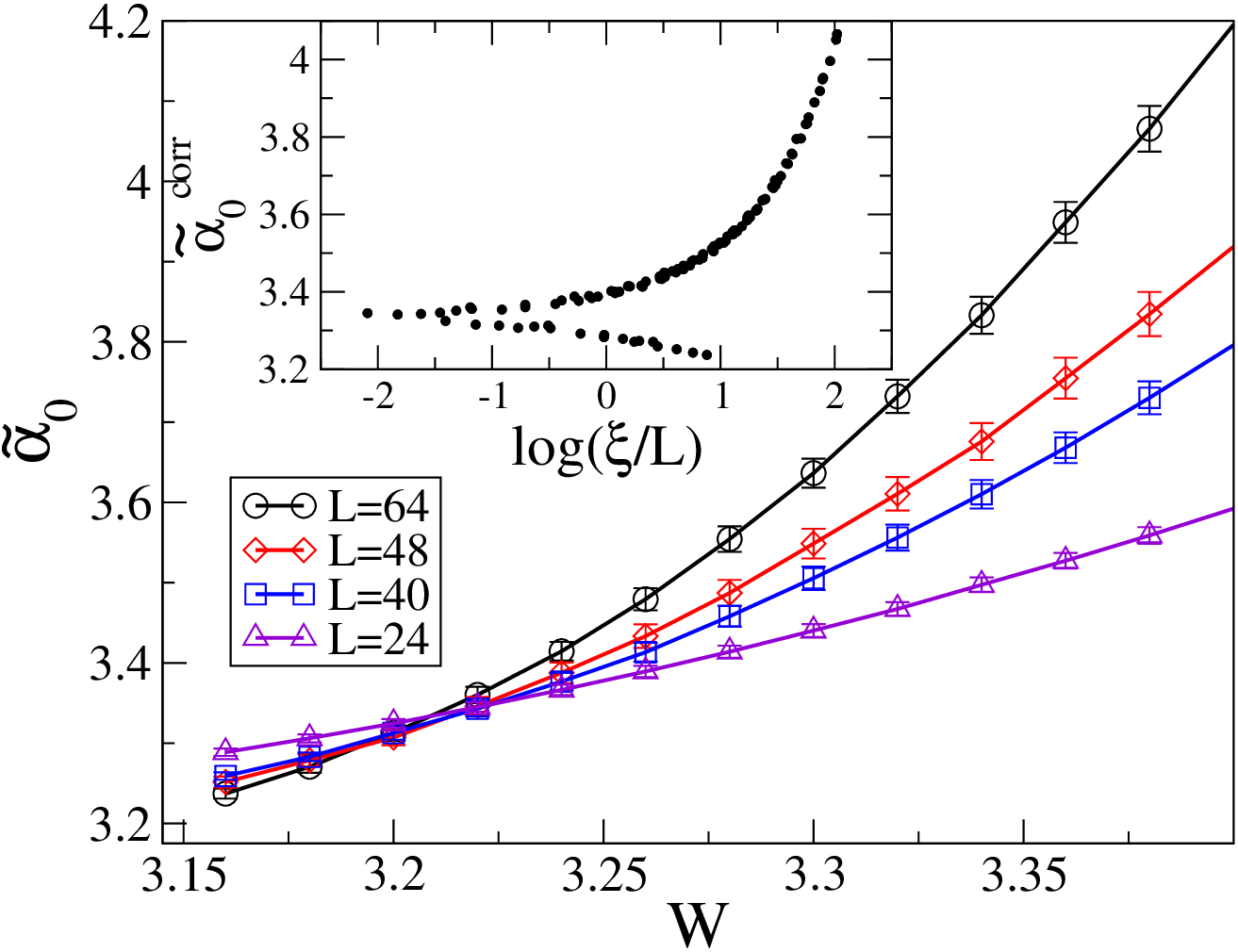}
 \caption{ The multifractal exponent $\alpha_0$ as a function of disorder strength $W$ that exhibits scaling 
   behavior around the critical disorder strength $W=3.2$. The inset shows the data collapse into  after performing the finite size scaling and plotting the scaling
   function for both branches of $\tilde{\alpha_{0}}$ in Eq. \ref{scalingCollapse}. The critical parameters
   used are $W_c=3.21$, $\nu=1.09$, $y=15.94$.  The orders of 
   expansion used for $G^{(0)}, G^{(1)}, \rho$, and $\eta$ are 
   $n_0=2, n_1=2, m_{\rho} = 1$ and  $m_{\eta}=0$ respectively. 
   The resulting $\chi^2 = 22$. The fit was chosen 
   by keeping the order of expansion low and taking the smallest $\chi^2$ for which the data collapse close to the fitting function  $\tilde{\alpha_{0}}$. }
 \label{fig:fssAlpha0}
\end{figure}

We can then calculate $\tilde{\alpha}_q$ for $q=0$  (given by Eq.\ref{alphaqForm} which we denote as $\tilde{\alpha}_0$) and is 
plotted in
Fig.\ref{fig:fssAlpha0}
as a function of system size and disorder strength which is expected to 
show the characteristic finite size behavior and exhibit a crossing
at the critical disorder strength\cite{Rodriguez2011}\cite{Rodriguez2010}.  We also carry out
multifractal finite size scaling for fixed $\lambda$ and
we assume our data $y_i$ (with uncertainty $\sigma_i$) 
is uncorrelated (as we only consider
fixed $\lambda$ so each point is from it's own realization) and thus the $\chi^2$
statistic for our model fits $f_i$ is
\beq
\chi^2 = \sum_i \frac{(y_i - f_i)^2}{\sigma_i^2}
\eeq
The order of expansion in $n_0$, $n_1$, $m_\eta$ and $m_\rho$ is determined by choosing the fit that keeps the $\chi^2$ statistic small, 
keeps the order of expansion small and provides a ``good'' 
collapse of the data into two branches.  Error bars
in fitting parameters are determined by generating new values of $\langle S_q \rangle$ and $\langle R_q \rangle$
for each corresponding $L$ and $W$ by pulling from a Gaussian distribution with 
mean $\langle S_q \rangle$ and variance 
$\sigma_{\langle S_q \rangle}/\sqrt{N-1}$ where N is the number of 
samples of $S_q$ and this is likewise done for $\langle R_q \rangle$.  This allows for a new
calculation of $\alpha_{q}$.  The result from this procedure
yields $Wc = 3.208 \pm 0.007$ and $\nu = 0.97 \pm 0.06$ in 
agreement with the above transfer matrix study.  All simulation parameters used for the calculation of the bogolon wave functions is collected in Sec.\ref{parameters}

\subsection{Discussion}
\label{sec:discussion}

It has been established by the work of  Ref. [\onlinecite{vish2001}] that the exponent $\nu$ is much 
different than the Anderson model.  We confirm this with our multifractal analysis,
establishing that this falls outside the Wigner-Dyson (WD) symmetry class.

The motivation for studying models of disordered superconductors is the 
rich variety of unusual properties they can exibit such as an enhanced single particle 
energy gap that persists even after superconductivity is destroyed \cite{Feigelman2007}.  Specific to this paper,
the motivation for studying the multifractal character of the eigenstates
is the proposal that multifractility can lead to 
enhancements of the critical temperature at which 
superconductivity is destroyed ($T_c$)\cite{Burmistrov2012,Mayoh2015}
which is observed in thin superconducting films that are weakly
disordered, namely Al\cite{adams2}\cite{Abeles1966} wich is still not well
understood.
An explanation for the enhancement of $T_c$ due to multifractility is that 
multifractility implies a broad distribution of exponents for the spatial
correlations at the transition (given by $f(\alpha)$).  This can be understood by the fact that there are regions of the system that have exponents that will decay off more slowly than if there were only a single one, implying stronger correlations among bogolon wavefunction $|\psi_i|^2$.  It is known that the regions of large $|\psi_i|^2$ for the
lowest excitations will correspond to regions of large local pairing amplitude $\Delta_i$\cite{nandiniBook} \cite{nandini2001}, and so $\Delta_i$
will also realize multifractal correlations.
The result of the longer range correlations would lead to stronger pairing 
correlations, resulting in an increase in $T_c$. Given the present calculations are done with a fixed distribution of $\Delta_{i}$, we cannot address quantitatively the relation between the $T_{c}$ and the disorder. 

Furthermore, it is known that the presence of bogolon excitations is 
what dissipates momentum and disrupts the flow of super current, destroying
superconductivity\cite{Lancaster2014}.  Therefore, a state in which the excitations are localized would help to ``protect'' superconductivity 
at finite temperatures and increase $T_c$.   
As the localization effect would be very strong in a 
quasi-2D system, when a superconducting film is made more thin the bogolons must become localized.  The reason it is not observed for all thin films 
(it is more typical for $T_c$ to decrease) is that if
the disorder is strong this effect will not be observed because strong disorder is already destroying the superconductivity as it destroys the long 
range phase coherence\cite{Liu1993}. 

Finally, we note that the multifractal analysis 
used here could be applied to models of 
conventional s-wave superconductivity with disorder 
which has been well studied \cite{Ghosal2001,Kamar2014,Jiang2013,Sakaida2013,Seibold2012,Sacepe2011,Bouadim2011,Aryanpour2007}.
This is important because the transfer matrix method cannot 
be used to locate the localization transition if the pairing 
must be solved self-consistently as this creates a correlation
between layers \cite{Qin2005}.  However, as all that is needed is the wavefunction
for this method, the multifractal finite size scaling analysis 
could be applied.

\section{Conclusion}\label{sec:conclusion}

We conclude that the multifractal analysis that works for the 
Anderson model can also be used 
for models of disordered superconductors to find the localization transition
of the quasi particle excitations.  In addition, it also confirms that 
the thermal metal to thermal insulator is indeed in a separate universality
class from the Anderson model. \cite{vish2001}  

Future work would include addressing the question of the relation between
multifractility of critical wavefunctions and the impact on $T_c$ more 
directly by finding the transition temperature for a  model of a conventional s-wave superconductor by solving the pairing field $\Delta_i$ 
self consistently for a given attraction interaction strength $U$. \cite{nandini2001}  
The multifractal spectrum $f(\alpha)$ could then be compared as a function of interaction strength and $T_c$ to quantitatively address the role played by multifractal eigenstates
and coupling strength on the critical temperature. Also, the question of whether this method can detect the superconductor to insulator transition \cite{Ghosal2001} would be of interest as this model could not be studied with transfer matrix due to the self consistency requirement on the pairing.

\section{Appendix A: Multifractal System Parameters}
\label{parameters}
In the TABLE~\ref{paramTable}, are the number of realizations (in the units of 1000 realizations) used for the
calculation of the bogolon wavefunctions where $L$ is the linear system size,
$W$ the disorder strength and $N_R$ the number of realizations.
In TABLE~\ref{chisqTable}, we test effects of the order of expansion of the fitting functions. The table lists the $\chi^2$ obtained as a function of $n_{0}$, $n_{1}$, $m_\rho$, and $m_\eta$. See Eq. \ref{Geq}, \ref{rhoeq}, and \ref{etaeq} for their definitions.

\begin{center}

\begin{table}[!htbp]
  \begin{tabular}{ | p{1cm} | p{1cm} | p{1cm} | p{1cm} | p{1cm}| p{1cm}| p{1cm} |}
    \hline
    \backslashbox{$W$}{$L$}  & $24$ & $32$ & $40$ & $48$ & $56$ & $64$ 
\\ \hline
        \hline
3.16  &  20 & 20 & 14 & 3.2 & 3.2 & 3.2 \\ \hline 
3.18  &  20 & 20 & 14 & 3.2 & 3.2 & 3.2 \\ \hline 
3.20  &  20 & 20 & 14 & 3.2 & 3.2 & 3.2 \\ \hline 
3.22  &  20 & 20 & 14 & 3.2 & 3.2 & 3.2 \\ \hline 
3.24  &  80 & 40 & 30 & 10  &  8  &   4\\ \hline 
3.26  &  80 & 40 & 30 & 10  &  8  &   4\\ \hline 
3.28  &  80 & 40 & 30 & 10  &  8  &   4\\ \hline 
3.30  &  80 & 40 & 30 & 10  &  8  &   4\\ \hline 
3.32  &  80 & 40 & 30 & 10  &  8  &   4\\ \hline 
3.34  &  80 & 40 & 30 & 10  &  8  &   4\\ \hline 
3.36  &  80 & 40 & 30 & 10  &  8  &   4\\ \hline 
3.38  &  80 & 40 & 30 & 10  &  8  &   4\\ \hline 
3.40  &  80 & 40 & 30 & 10  &  0  &   0\\ \hline 
\end{tabular}
\caption{Number of realizations as a function of $W$ and $L$. The number of realizations is in the units of 1000 realizations. }
\label{paramTable}
\end{table}
\end{center}

\begin{center}
\begin{table}[!htbp]

 \begin{tabular}{ | p{1cm} | p{1cm} | p{1cm} | p{1cm}| p{2cm} |}
    \hline
    $n_{0}$  & $n_{1}$ & $m_\rho$ & $m_\eta$ &  $\chi^2$  \\ \hline
    \hline
2  &  0  &  1 &  0 & 15.55 \\ \hline
 2  &  1  &  1 &  0 & 15.55 \\ \hline
 3  &  0  &  1 &  0 & 15.20 \\ \hline
 3  &  1  &  1 &  0 & 15.20 \\ \hline
 2  &  0  &  1 &  1 & 15.55 \\ \hline
 2  &  1  &  1 &  1 & 15.55 \\ \hline
 3  &  0  &  1 &  1 & 15.20 \\ \hline
 3  &  1  &  1 &  1 & 15.20 \\ \hline
 2  &  0  &  2 &  0 & 15.54 \\ \hline
 2  &  1  &  2 &  0 & 15.54 \\ \hline
 3  &  0  &  2 &  0 & 14.67 \\ \hline
 3  &  1  &  2 &  0 & 14.67 \\ \hline
 2  &  0  &  2 &  1 & 15.54 \\ \hline
 2  &  1  &  2 &  1 & 15.54 \\ \hline
 3  &  0  &  2 &  1 & 14.67 \\ \hline
 3  &  1  &  2 &  1 & 14.67 \\ \hline
 2  &  0  &  2 &  2 & 15.54 \\ \hline
 2  &  1  &  2 &  2 & 15.54 \\ \hline
 3  &  0  &  2 &  2 & 14.67 \\ \hline
 3  &  1  &  2 &  2 & 14.67 \\ \hline
 2  &  0  &  3 &  0 & 15.21 \\ \hline
 2  &  1  &  3 &  0 & 15.21 \\ \hline
 3  &  0  &  3 &  0 & 14.63 \\ \hline
 3  &  1  &  3 &  0 & 14.63 \\ \hline
 2  &  0  &  3 &  1 & 15.21 \\ \hline
 2  &  1  &  3 &  1 & 15.21 \\ \hline
 3  &  0  &  3 &  1 & 14.63 \\ \hline
 3  &  1  &  3 &  1 & 14.63 \\ \hline
 2  &  0  &  3 &  2 & 15.21 \\ \hline
 2  &  1  &  3 &  2 & 15.21 \\ \hline
 3  &  0  &  3 &  2 & 14.63 \\ \hline
 3  &  1  &  3 &  2 & 14.63 \\ \hline
\end{tabular}

  \caption{$\chi^2$ dependence on order of expansion}
    \label{chisqTable}
\end{table}
\end{center}

\begin{acknowledgements}
This work is supported by NSF EPSCoR Cooperative Agreement No. EPS-1003897 (C.W.M., K.-M.T., Y.Z., and M.J.). This work  used  the  high  performance  computational resources provided by the Louisiana Optical Network Initiative (http://www.loni.org) and HPC@LSU computing.  Additional support (MJ) was provided by NSF Materials Theory grant DMR1728457.
\end{acknowledgements}

\end{document}